\documentclass[]{aastex}
\usepackage{emulateapj5}
\usepackage{graphicx}
\newcommand{\myemail}{manojendu@ncra.tifr.res.in}
\shorttitle{Hard X-ray time lag}
\shortauthors{Choudhury et al.}
\submitted{Received 2004 August 17; Accepted 2004 October 07}

\begin{document}

\title{Detection of anti-correlated hard X-ray time lag in Cygnus X-3}
\author{Manojendu Choudhury\altaffilmark{1} \& A. R. Rao
}
\affil{Tata Institute of Fundamental Research, Mumbai-400005, India}
\altaffiltext{1}{Current address: N.C.R.A., T.I.F.R.,Pune-411007. India. \\ e-mail:
\myemail}

\begin{abstract} The wide-band X-ray spectra of the high mass X-ray binary Cygnus
X-3 exhibits a pivoting behavior in the `low' (as well as `hard') state, correlated
to the radio emission. The time scale of the soft and hard X-rays' anti-correlation,
which gave rise to the pivoting feature, was found to be less than a day from the
monitoring observations by {\it RXTE}--ASM and {\it CGRO}--BATSE. In this {\it
Letter} we report the detection of a lag of $\lesssim$ 1000s in the anti-correlation
of the hard X-ray emission (20--50 keV) to that of the soft X-ray emission
(2--7 keV), which may be attributed to the viscous time scale of flow of matter in
the accretion disk. This suggests the geometrical picture of a truncated accretion
disc with a Compton cloud inside the disc, the relative sizes of which determine the
spectral shape. Any change in the disc structure will take place in a viscous time
scale, with corresponding anti-correlated change in the Compton cloud. We also report
the pivoting in the spectra in one span of a pointed observation when an episode of
the rearranging of the accretion system is serendipitously observed. This is the
first such observation of hard X-ray delay seen in the persistent Galactic
microquasars, within the precincts of the hard state.

\end{abstract}


\section {Introduction}
The presence of a   non-thermal component in the wide-band X-ray spectrum is one of
the most prominent observational features of the Galactic blackhole candidates
as well as neutron star binaries \citep[see][for a short review]{barret04}, although the very precise nature
of its origin has not yet been definitively ascertained. Conflicting views regarding
the physical mechanism of the generation of hard X-ray emission, ranging from
synchrotron at the base of the jet \citep{markoff01aa, markoff03aa} to Comptonization
of thermal photons by a hot corona with various geometrical structures
\citep[see, for eg.,][]{poutanen98} -- the most favored being that of a hot
quasi-spherical cloud inside a truncated disc \citep{zdziarski02apj} -- exist in the
literature. \citet{mcclintock04} establish the various classes of X-ray emission
states, based on the spectral characteristics, adding to the canonical `low-hard'
and `high-soft' states \citep{tanaka95}. Of these, the `low-hard' state is perhaps
the most interesting for all blackhole sources for the presence of ubiquitous outflow
in the form of a collimated jet \citep{fender01mnras}. These jets are also present in
the low magnetic field ($\lesssim 10^9$ G) neutron stars \citep{kuulkers01mnras},
collectively these objects are classified as {\it microquasars}
\citep{mirabel99araa}. The ADAF \citep{narayan94apjl} and TCAF
\citep{chakrabarti96phr} models provide viable alternative explanations of the
accretion-ejection phenomena \citep[respectively]{meier01apjl, das99cqg} in this
class of sources, involving a truncated accretion disc.

Cygnus X-3 is an enigmatic X-ray binary system which is one of the brightest
Galactic sources in the radio band. The X-ray (as well as the infra-red) emission
exhibit a binary modulation of 4.8 hours \citep{singh02aa}. Although the nature of
the compact object is not yet known, the persistent radio emission from a jet
classifies it as a prominent microquasar, and the different X-ray spectral states of
this source \citep{choudhury02japa} are similar in nature to that of the canonical
blackhole candidates, although individual components describing the spectral shape
may be different \citep{choudhury03apj}. The long term statistical correlation among
the soft X-ray (2--12 keV, {\it RXTE}--ASM), hard X-ray (20--100 keV,
{\it CGRO}--BATSE) and radio emission (2.2 GHz, {\it GBI}) revealed a
pivoting behavior of the X-ray spectra (in the region of 10--20 keV) correlated to
the radio emission, within the precincts of the hard state of this source
\citep{choudhury02aa}. This behavioral feature, in
the `low-hard' state, has been established as a general feature of the X-ray emission
for the Galactic blackhole binaries, with the pivot point varying for the various
sources \citep{choudhury03apj}. In \citet{choudhury04aa} it has been established,
from the daily monitoring of the source simultaneously in the soft X-ray, hard X-ray
and radio, that
the correlation time scale for Cygnus X-3 is less than a day. Considering the
universal correlation of the soft X-ray flux with the
radio spanning 5 orders of magnitude of intrinsic luminosity, \citet{choudhury03apj}
postulated that it is the soft X-ray that determines the spectral pivoting phenomena which is the driving force of the evolution of the accretion-ejection mechanism in
these systems. In such a scenario, the Compton scattered X-rays originate from a
region close to the compact object, which may be confined within the Centrifugal
Boundary Layer (CENBOL) \citep{chakrabarti95apj, ebisawa96pasj}. At low accretion
rates, the disk is truncated far away from the compact object and the X-ray spectrum
is dominated by a thermal-Compton spectrum, originating from the high temperature
region within. Increase/decrease in accretion rate causes the location of the disk
truncation (CENBOL) to move in/out, causing the Comptonizing cloud to
decrease/increase as well as the amount of seed thermal photon to increase/decrease
which cools the Comptonizing cloud more/less efficiently, giving rise to the pivoting
of the wide-band X-ray spectrum. Hence, the pivoting of the X-ray spectrum is not
expected to occur instantaneously, as the viscous time scale of the flow of matter in
the accretion disc determines the change in the hard X-rays vis-a-vis the soft
X-rays, and in such a situation the delay of the hard X-ray should be much less than
a day \citep{choudhury04aa}, implying the need to search for delay between the hard
and soft X-rays using the long pointed observations, spanning at least a good
fraction of one binary orbit. Considering the fact that the mechanism of a change in
accretion rate is very unlikely to be a continuous process, the individual events of
such a change in the soft X-ray emission with an eventual anti-correlated change in
the hard X-ray emission need to be serendipitously observed during the pointed
observations, with the source in the `low' (as well as `hard') state.

\vskip 0.3cm
{
\includegraphics[width=7cm,height=4cm,angle=0]{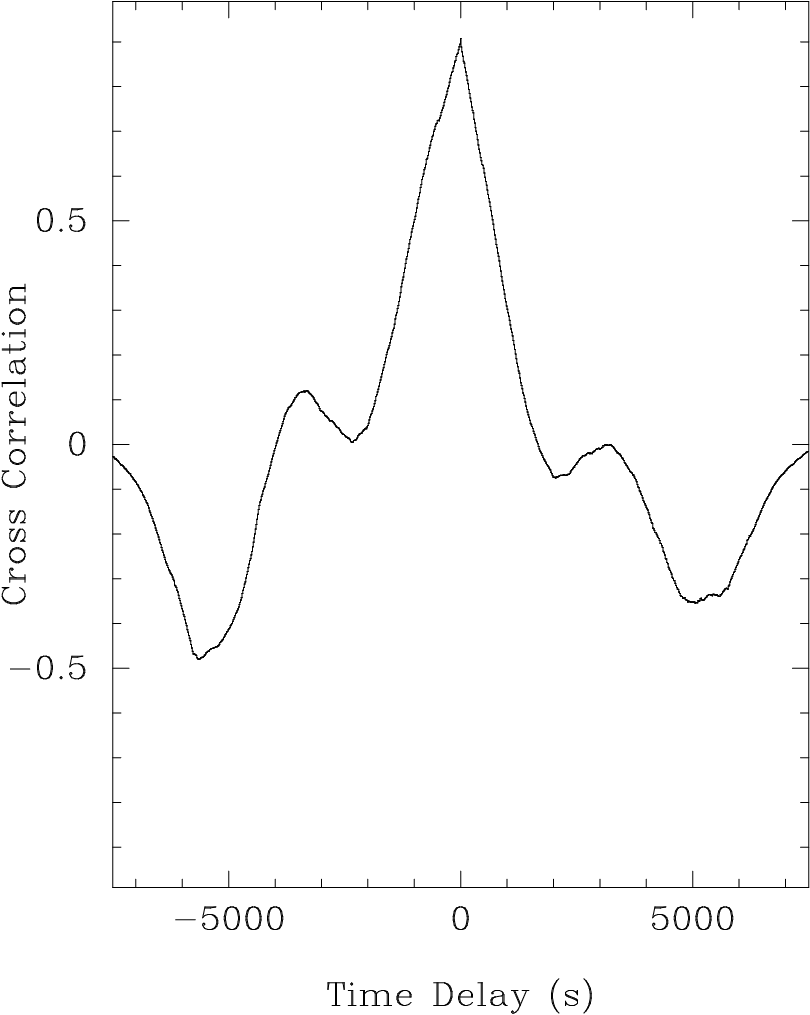} \\
{{\bf Figure 1.} A typical cross-correlation between the soft (2-7 keV) and hard (20-50 keV) X-ray flux with the lightcurve not corrected for binary modulation. This particular result was obtained for observations on MJD 50953.\label{fig1}}
}
\vskip 0.2cm

In this {\it Letter}, we report the detection of such an anti-correlated delay, of
$\lesssim$ 1000 s of the hard X-rays (20--50 keV) with respect to the soft X-ray flux
(2--7 keV)., i.e. any change in the soft X-ray flux results in an opposite change in
the hard X-ray flux after the observed delay time scale, during the individual
pointed observations using the PCA and HEXTE aboard the {\it RXTE}, in the `low'
(correspondingly `hard') state of the X-ray emission in the source. This type of
lagged anti-correlation, we believe, is the first to be observed for the class
of X-ray persistent Galactic microquasars within the bounds of the (low-) hard state.
We confirm our detection of this anti-correlated delay by observing the pivoting of
the spectra `in-situ' during one of the pointed observations of the {\it RXTE}--PCA.

\section {Data and Analysis}
The data are obtained from the archives of the {\it RXTE} through the HEASARC Online
Service. The pointed observations of both the narrow field
of view instruments aboard the {\it RXTE}, viz. PCA and HEXTE (cluster 0) are used
for initial study of the cross-correlation of the lightcurves at the various energy
bands, whereas the results of binary corrected lightcurves are reported here only for
the PCA observations. The lightcurves as well as the spectrum (obtained from the PCA)
uses the standard 2 form of data (all PCUs added), with all the procedures of data
filtering, background and deadtime corrections strictly adhered to. The basic data
reduction and analysis was carried out using HEASOFT (v5.2). The correction for binary
modulation was done using the recipe described by \citet{choudhury04aa}, using the
binary template of \citet{vanderklis89aa} and obtaining the phase of each pointing
from the quadratic ephemeris of \citet{singh02aa}.

\vskip 0.3cm
{
\includegraphics[width=7cm,height=8cm,angle=0]{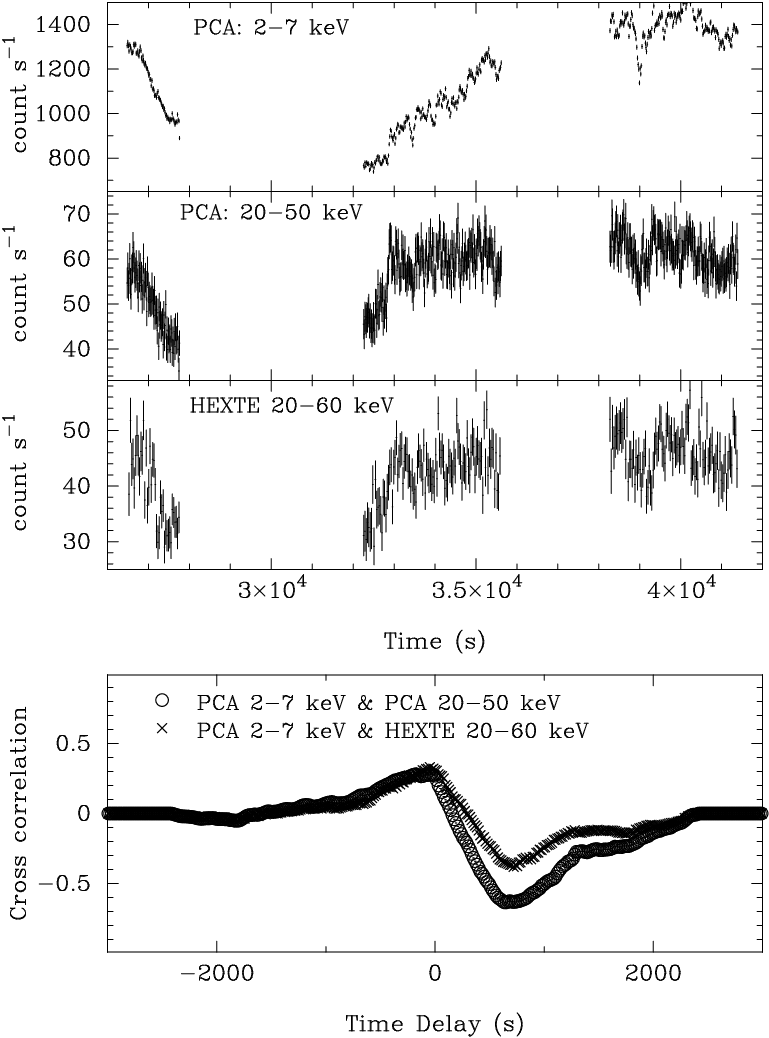} \\
{{\bf Figure 2.} The (binary uncorrected) lightcurve of the source on MJD 50321 as observed by the PCA at 2-7 keV (top panel) and 20-50 keV (middle panel), along with the HEXTE at 20-60 keV (bottom panel). The cross-correlation shows a clear indication of possible anti-correlated delay of the hard X-ray flux.\label{figc}}
}
\vskip 0.2cm

The pointed observations of this source from the {\it RXTE} were done predominantly
in the `high' state, with the emphasis on the investigation during the massive radio
flaring episodes. As a consequence, only a few observations in the `low' state exist
for this source and that too during only two phases, MJD 50319--50325 and
50949--50954. During the first phase the electron background was high and one needs
to filter the data very judiciously, while on the other occasion the source exhibited its `hardest'
state, providing one extreme of the X-ray spectral characteristics. The binary
modulation of the X-ray emission is very strong in the source. In the `low' state,
the soft X-ray flux (2--7 keV) is modulated by a factor $\sim$2 and the hard X-ray
flux (20--50 keV) is modulated by a factor $\sim$2.5--3. This strong modulation
smothers all other variations in nearly all the cases, and hence the
cross-correlation shows a strong correlation between the emission at the two flux
ranges (Figure \ref{fig1}). Extensive search for anti-correlated delay from the
binary modulated lightcurves yielded only one possible indication of the presence
of such a delay, on MJD 50321 (obs. id.: 10126-01-01-01). The strong anti-correlation
with a delay of $\sim$500-700 s is seen for hard X-rays from both PCA (20--50 keV)
and HEXTE (20--60 keV) with respect to the soft X-rays (2--7 keV, PCA), as
illustrated in Figure \ref{figc}. The positive correlation without any delay, due to
the binary modulation is as prominent (F.F.T. correlation coefficient - hereafter CC
- of 0.3 for 488 degrees of freedom - hereafter dof, Pearson's CC of 0.56 for 244 dof
- with a null probability $\sim$0, and Spearman's CC of 0.54 with 488 dof | null
probability $\sim$0) as the negative correlation with a delay of a few hundred
seconds (CC of -0.6 for the PCA data). When the smooth and strong binary variation is
subtracted using the binary ephemeris \citep{choudhury04aa}, the positive correlation
got suppressed and the negative correlation with a delay  was the only prominent
feature of the cross-correlation plot.

\vskip 0.3cm
{
\includegraphics[width=8.5cm,height=8.5cm,angle=0]{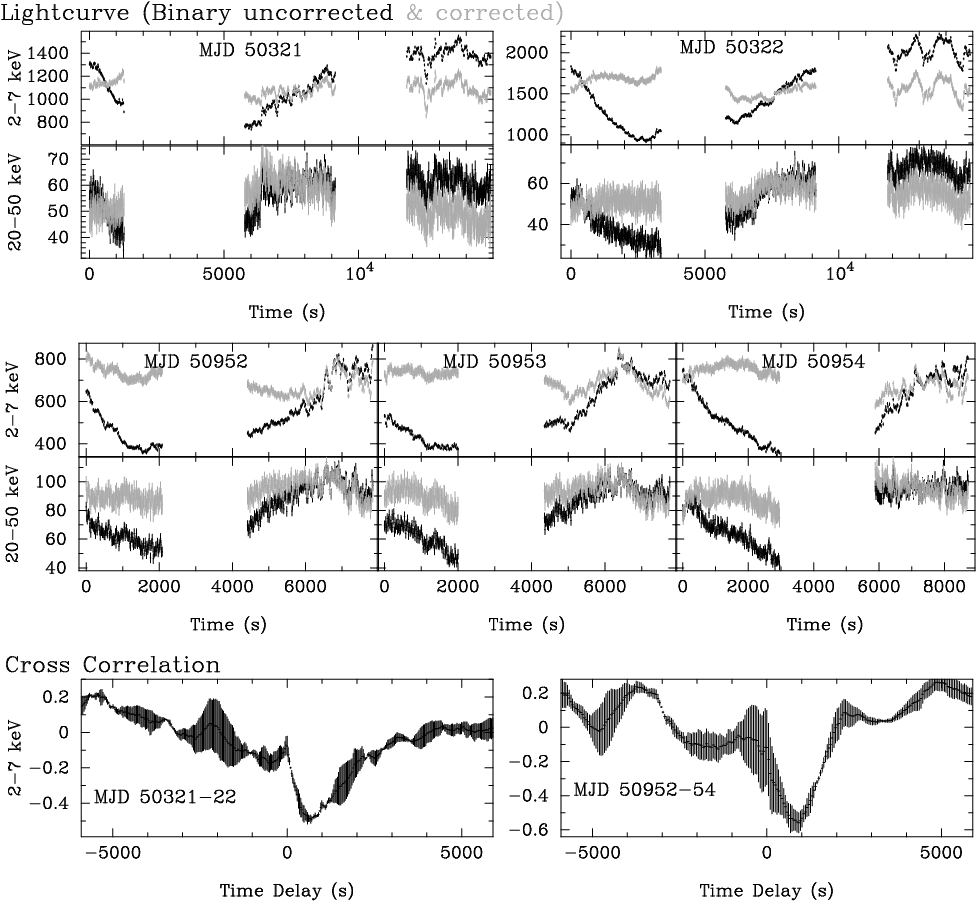} \\
{{\bf Figure 3.} The lightcurve as observed by the PCA at 2-7 keV and 20-50 keV, both uncorrected and corrected for the binary modulation, on the five days showing the anti-correlated hard X-ray lags. The cross-correlation plots sown are obtained after co-adding the lightcurves of MJD 50321 \& 50322, and MJD 50952, 50953 \& 50954 (two phases), respectively, to compute the maximum possible error in the measurement. The details of the individual pointed observations are given in table \ref{tab1}.\label{figd}}
}
\vskip 0.2cm

Repeating the cross-correlation test for these `low' (and `hard') states after
correcting for the binary modulation \citep{choudhury04aa} provided very significant
results, showing a very strong anti-correlated delay of the hard X-rays (20--50 keV,
PCA) with that of the soft X-rays (2-7 keV, PCA), on five days, including MJD 50321.
Table \ref{tab1} gives the details of the days of observation of the delay, the
observation id and the time delay observed on the particular pointings. The delay is
first obtained by the F.F.T. algorithm using the XRONOS program "crosscor", the
Pearson's and Spearman's coefficients are subsequently obtained. The F.F.T.
coefficient is normalized by (dividing by) the square root of the product of number
of good newbins of the concerned lightcurves, effectively this coefficient is the
cross covariance of the two lightcurves. To substantiate the statistical robustness
of the observed delay we co-add the lightcurves (in two phases, MJD 50321-22 \&
50952--54) to obtain the maximum possible error in the correlation coefficient. The
cross-correlation of the co-added lightcurves, along with the individual lightcurves
(both binary uncorrected and corrected) are presented in Figure 3, where the
delay as well as cross-correlation is established, despite the intrinsic variation in
the delay as well as the correlation coefficient of the individual pointings (table
\ref{tab1}). 

The spectral evolution of the source during the pointed observation on MJD 50321
(obs. id.: 10126-01-01-01) is shown in Figure 4, where the two spectra
correspond to the softer and harder regions of the lightcurve, i.e. before and after
pivoting. The pivoting of the spectra, earlier reported for observations taking place
on different days corresponding to the extreme cases within the bounds of the `low'
and `hard' states \citep{choudhury02aa, choudhury03apj}, is observed directly within
the span of one pointed observation containing the episode of the soft X-ray flux
driving the spectral evolution, and the hard X-ray flux changing correspondingly
after a time scale of the viscous flow in the accreting fluid. This `in situ'
observation of pivoting, in the hard state, validates the cross-correlation results
and puts the particular observational feature on very firm footing.

\section{Discussion}

The strong binary modulation of Cygnus X-3 has prevented any serious study on the
variability of this source at times-scales of few hundred seconds to about a day.
\citet{singh02aa} have shown that the binary period (and the template of its
variation) has remained consistent for more than 25 years  with a linear decay with
no second derivative of the period. \citet{choudhury04aa} have used this information
and shown that this ephemeris can be used to remove the large amplitude smooth binary
variation without recourse to any derivation of phase during individual observations
- the only free parameters being the amplitude of variation and the mean count rate.
Using this method, we have corrected for the large amplitude binary variations and
detected anti-correlated hard X-ray delays in five of the observations.

The only assumption incorporated in our method is that the occasional jitters seen in
the binary modulation is not due to any binary instability \citep{choudhury04aa},
rather we explore whether any residual energy dependent phase jitter and/or variable
energy dependent template could have caused the anti-correlated delay presented here.
In all observations we find a strong positive correlation with zero phase indicating
the similarity of the template in the two energy bands. Further, more than 90\% of
the binary variability in Cygnus X-3 is energy independent \citep[see][]{rajeev94apj}
and the large amplitude variations in high phase are correlated in all energies. The
anti-correlated behavior is seen in the slow varying components which is unrelated
to the binary variation. This was evident when we examined the variation of the
hardness ratio compared to the lightcurves.

The strongest evidence for the reality of the anti-correlation comes from the model
independent spectral pivoting seen in one of the observations (see Fig 4).
\citet{choudhury03apj, choudhury02aa} have shown that this type of spectral pivoting
at 10--20 keV energies occurs when one takes spectrum at different days at different
count rates. The fact that such spectral pivoting is indeed seen in one single
observation shows that the soft and hard X-ray anti-correlation occurs in a time
scale less than the binary period and the cross-correlation analysis reveals the
delay in the time scale close to a thousand seconds.

\vskip 0.3cm
{
\includegraphics[width=7cm,height=4cm,angle=0]{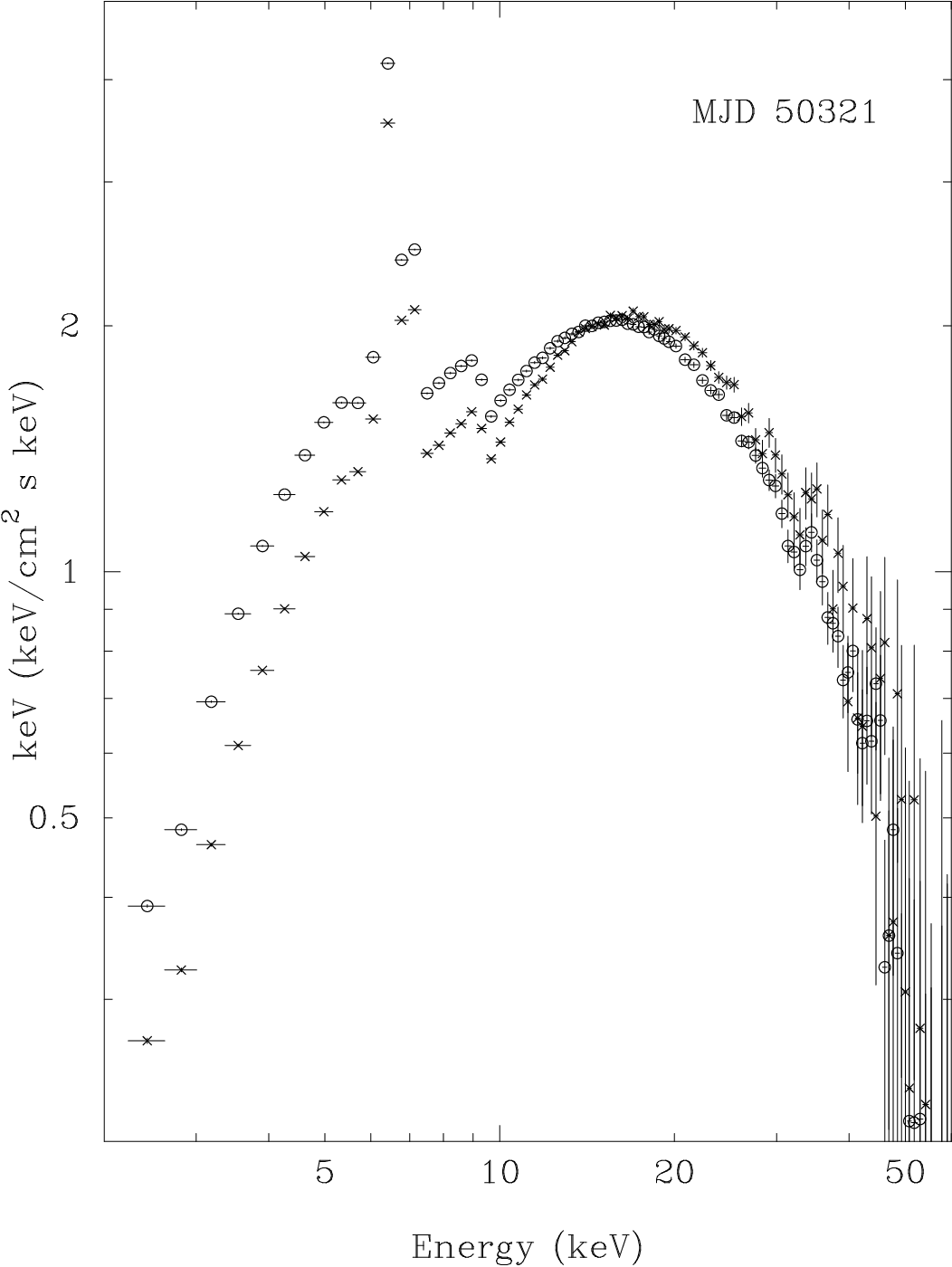} \\
{{\bf Figure 4.} The wide-band X-ray spectra of the source on MJD 50321, for the soft and the hard regions of the lightcurve, resulting in comparative softer and harder spectral distribution, displaying the `in situ' pivoting in the region of 10--20 keV.\label{fig4}}
}
\vskip 0.2cm

The three-way association between soft X-rays, hard X-rays and radio emission in 
Cygnus X-3 maybe explained by the geometry of truncated, optically thick, accretion
disc, with a Comptonizing cloud of energetic electrons inside which may have
advective \citep{narayan94apjl} or bulk motion \citep{chakrabarti95apj}. The soft
X-ray flux determines the accretion disk structure, while the Compton cloud within
the truncated disc is responsible for the hard X-ray emission and it also drives the
outflow \citep[see][for contrasting paradigms]{meier01apjl, das99cqg}. The delay
presented in this work signifies the disk readjustment time scale which should
correspond to the location of the truncation radius. The viscous time $t_{vis}$ can
be calculated  for the radiation pressure dominated optically thick accretion disc as
$$t_{vis} = 30 \alpha^{-1} M^{-1/2} R^{7/2} \dot{M}^{-2} s$$ where $\alpha$ is the
viscosity parameter in units of 0.01, M is the mass of the compact object in solar
mass units, R is the radial location in the accretion disc in units of 10$^7$ cm,
$\dot{M}$ is the mass accretion rate in units of 10$^{18}$ g/s
\citep[see][]{belloni97apj_b,fender04araa}. Taking $\alpha$ = 1, M = 10 and
$\dot{M}$ = 3 (corresponding to one fifth of the Eddington accretion rate), we get R
$\sim$ 7 for a viscous time scale of 1000 s. Thus, if Cyg X-3 harbors a black hole of
10 M$_\odot$, the observed delay signifies a CENBOL location
of $\sim$ 25 Schwartzchild radius.

It should be noted that the X-ray emission shows a strong correlation with the
radio emission, but the radio emission does not show any binary modulation.
It would be interesting to repeat the type of work presented here with a
simultaneous radio monitoring observations so that a clear handle can be
obtained for the location of the radio emission.

\section*{Acknowledgments} This research has made use of data obtained through the 
HEASARC Online Service, provided by the NASA/GSFC, in support of NASA High Energy
Astrophysics Programs.

\vskip -0.3cm
\begin{deluxetable}{llclll}
\tabletypesize{\scriptsize}
\tablecolumns{6}
\tablewidth{0pc}
\tablecaption{The details of the pointed observations for which the lagged anti-correlation of the hard X-rays (20--50 keV, PCA) with respect to the soft X-rays (2--7 keV,
PCA) has been found, after correcting for the binary modulation, as prescribed by \citet{choudhury04aa}. The details of the delay observed in the individual lightcurves observed on the particularly mentioned days is given.\label{tab1}}
\tablehead{
\colhead{MJD} & \colhead{Observ. Id.} & \colhead{Delay (s)} & \multicolumn{3}{c}{Statistical coefficient} \\
\cline{4-6}
\colhead{} & \colhead{} & \colhead{(error)} & \colhead{F.F.T. (error)} & \colhead{Pearson (Null Prob.)} & \colhead{Spearman (Null Prob.)} 
}
\startdata
50321&10126-01-01-01& $\sim$ 620 ($\pm70$) & $\sim$ -0.48 ($\sim0.04$) & -0.51 ($\sim10^{-12}$) & -0.58 ($\sim10^{-30}$) \\
50322&10126-01-01-02& $\sim$ 750 ($\pm120$)& $\sim$ -0.44 ($\sim0.04$) & -0.46 ($\sim10^{-4}$) & -0.43 ($\sim10^{-13}$) \\
50952&30082-04-04-00& $\sim$ 700($\pm50$) & $\sim$ -0.61 ($\sim0.08$) & -0.66 ($\sim10^{-11}$) & -0.69 ($\sim10^{-36}$) \\
50953&30082-04-05-00& $\sim$ 950($\pm60$) & $\sim$ -0.58 ($\sim0.08$) & -0.73 ($\sim10^{-13}$) & -0.75 ($\sim10^{-39}$) \\
50954&30082-04-06-00& $\sim$ 0--1000  & $\sim$ -0.40 ($\sim0.08$) & -- & -- \\
\enddata
\end{deluxetable}


\begin{thebibliography}{}
\bibitem[Barret (2004)]{barret04} Barret, D. 2004, proc. of Plasmas in the Laboratory and in the Universe, American Institute of Physics Conference Series, 238-249 [astro-ph/0401100]
\bibitem[Belloni et al.(1997)]{belloni97apj_b} Belloni, T., Mendez, M., King, A. R., van der Klis, M. \& van Paradijs, J. 1997, \apj, 488, L109 
\bibitem[Chakrabarti (1996)]{chakrabarti96phr} Chakrabarti, S.K. 1996, \physrep, 266, 229
\bibitem[Chakrabarti \& Titarchuk (1995)]{chakrabarti95apj} Chakrabarti, S. K., \& Titarchuk, L. G. 1995, \apj, 455, 623
\bibitem[Choudhury \& Rao (2002)]{choudhury02japa} Choudhury, M., \& Rao, A.R. 2002, JApA, 23, 39
\bibitem[Choudhury et al. (2004)]{choudhury04aa} Choudhury, M., Rao, A.R., Vadawale, S.V., Jain, A.K. \& Singh, N.S. 2004, \aap, 420, 665
\bibitem[Choudhury et al. (2003)]{choudhury03apj} Choudhury, M., Rao, A.R., Vadawale, S.V. and Jain, A.K. 2003, \apj, 593,452
\bibitem[Choudhury et al. (2002)]{choudhury02aa} Choudhury, M., Rao, A.R., Vadawale, S.V., Ishwara-Chandra, C.H. \& Jain, A.K., 2002, \aap, 383, L35
\bibitem[Das \& Chakrabarti (1999)]{das99cqg} Das, T. \& Chakrabarti, S.K. 1999, Class \& Quant. Gravity, 16, 3879
\bibitem[Ebisawa et al. (1996)]{ebisawa96pasj} Ebisawa, K., Titarchuk, L. \& Chakrabarti, S. 1996, \pasj, 48, 59
\bibitem[Fender \& Belloni (2004)]{fender04araa} Fender, R.P. \& Belloni, T. 2004, \araa, 42, 317-364
\bibitem[Fender (2001)]{fender01mnras} Fender, R.P. 2001, \mnras, 322, 31
\bibitem[Fender \& Kuulkers (2001)]{kuulkers01mnras} Fender, R.P. \& Kuulkers, E. 2001, \mnras, 324, 923

\bibitem[Markoff et al. (2001)]{markoff01aa} Markoff, S., Falcke, H. \& Fender, R. 2001, \aap, 372, L25
\bibitem[Markoff et al. (2003)]{markoff03aa} Markoff, S., Nowak, M., Corbel, S., et al. 2003, \aap, 397, 645
\bibitem[McClintock \& Remillard (2004)]{mcclintock04} McClintock, J.E. \& Remillard, R.A. 2004, to appear in {\it Compact Stellar X-ray Sources}, W.H.G. Lewin and M. van der Klis eds., Cambridge University Press, preprint (astro-ph/0306213)
\bibitem[Meier (2001)]{meier01apjl} Meier, D.L. 2001, \apj, 548, L9
\bibitem[Mirabel \& Rodriguez (1999)]{mirabel99araa} Mirabel, I.F. \& Rodriguez, L.F. 1999, \araa, 37, 409
\bibitem[Narayan \& Yi (1994)]{narayan94apjl} Narayan, R. \& Yi, I. 1994, \apj, 428, L13
\bibitem[Poutanen (1998)]{poutanen98} Poutanen, J.  1998, in {\it Theory of Black Hole Accretion Discs},  M.~A. Abramowicz, G. Bjornsson \& J.~E. Pringle, eds.,  Cambridge Contemporary Astrophysics, Cambridge University Press, 100p.
\bibitem[Rajeev et al. (1994)]{rajeev94apj} Rajeev, M.R., Chitnis, V.R., Rao, A.R., et al. 1994, \apj, 424, 376
\bibitem[Singh et al. (2002)]{singh02aa} Singh, N.S., Naik, S., Paul, B. et al. 2002, \aap, 392,161
\bibitem[Tanaka \& Lewin (1995)]{tanaka95} Tanaka, Y. \& Lewin, W.H.G. 1995, Cambridge Astrophys. Ser. vol. 26, X-Ray Binaries, eds. W.H.G. Lewin, J. van Paradijs \& E.P.J. van den Heuvel (Cambridge Univ. Press), 126
\bibitem[van der Klis \& Bonnet-Bidaud (1989)]{vanderklis89aa} van der Klis, M. \& Bonnet-Bidaud, J.M. 1989, A\&A, 214, 203
\bibitem[Zdziarski et al. (2002)]{zdziarski02apj} Zdziarski, A.A., Poutanen, J., Paciesas, W.A., et al. 2002, \apj, 578, 357

\end{thebibliography}
\end{document}